\documentclass[aps,prl,twocolumn,superscriptaddress,showpacs,10pt]{revtex4-1}

\usepackage{amsmath, amsthm, amssymb}
\usepackage{bm}
\usepackage{braket}
\usepackage{array}
\usepackage{longtable}
\usepackage{color}  
\usepackage{graphicx}
\usepackage{dcolumn}
\newcolumntype{.}{D{.}{.}{-1}}
\usepackage{multirow}
\usepackage[figuresright]{rotating}
\usepackage{units}
\usepackage{subfigure}
\usepackage[latin1]{inputenc}
\usepackage{textcomp}
\usepackage{gensymb}
\usepackage[rightcaption]{sidecap}
\usepackage{nicefrac}
\usepackage{xspace}
\usepackage[version=3]{mhchem}

\newcommand{\cacro}{Ca$_{10}$Cr$_7$O$_{28}$\xspace}

\newcommand{\Cr}{Cr$^{5+}$\xspace}

\begin{document}

\title{The magnetic Hamiltonian and phase diagram of the quantum spin liquid \cacro}

\author{Christian Balz}
\email{christian.balz@helmholtz-berlin.de}
\affiliation{Helmholtz-Zentrum Berlin f\"ur Materialien und Energie, 14109 Berlin, Germany}
\affiliation{Institut f\"ur Festk\"orperphysik, Technische Universit\"at Berlin, 10623 Berlin, Germany}
\author{Bella Lake}
\affiliation{Helmholtz-Zentrum Berlin f\"ur Materialien und Energie, 14109 Berlin, Germany}
\affiliation{Institut f\"ur Festk\"orperphysik, Technische Universit\"at Berlin, 10623 Berlin, Germany}
\author{A.T.M. Nazmul Islam}
\affiliation{Helmholtz-Zentrum Berlin f\"ur Materialien und Energie, 14109 Berlin, Germany}
\author{Yogesh Singh}
\affiliation{Indian Institute of Science Education and Research (IISER) Mohali, Knowledge City, Sector 81, Mohali 140306, India}
\author{Jose A. Rodriguez-Rivera}
\affiliation{NIST Center for Neutron Research, National Institute of Standards and Technology, 20899 Gaithersburg, USA}
\affiliation{Department of Materials Science, University of Maryland, College Park, 20742 Maryland, USA}
\author{Tatiana Guidi}
\affiliation{ISIS Facility, STFC Rutherford Appleton Laboratory, Oxfordshire OX11 0QX, UK}
\author{Elisa M. Wheeler}
\affiliation{Institut Laue-Langevin, 38042 Grenoble, France}
\author{Giovanna G. Simeoni}
\affiliation{Heinz Maier-Leibnitz Zentrum, Technische Universitat M\"unchen, 85748 Garching, Germany}
\author{Hanjo Ryll}
\affiliation{Helmholtz-Zentrum Berlin f\"ur Materialien und Energie, 14109 Berlin, Germany}

\date{\today}

\begin{abstract}
A spin liquid is a new state of matter with topological order where the spin moments continue to fluctuate coherently down to the lowest temperatures rather than develop static long-range magnetic order as found in conventional magnets. For spin liquid behavior to arise in a material the magnetic Hamiltonian must be "frustrated" where the combination of lattice geometry, interactions and anisotropies gives rise to competing spin arrangements in the ground state. Theoretical Hamiltonians which produce spin liquids are spin ice, the Kitaev honeycomb model and the Heisenberg kagome antiferromagnet. Spin liquid behavior however in real materials is rare because they can only approximate these Hamiltonians and often have weak higher order terms that destroy the spin liquid state. \cacro is a new quantum spin liquid with magnetic \Cr ions that possess quantum spin number S=\nicefrac{1}{2}. The spins are entirely dynamic in the ground state and the excitation spectrum is broad and diffuse as is typical of spinons which are the excitations of a spin liquid. In this paper we determine the Hamiltonian of \cacro using inelastic neutron scattering under high magnetic field to induce a ferromagnetic ground state and spin-wave excitations that can be fitted to extract the interactions. We further explore the phase diagram by using inelastic neutron scattering and heat capacity measurements and establish the boundaries of the spin liquid phase as a function of magnetic field and temperature. Our results show that \cacro consists of distorted kagome bilayers with several isotropic  ferromagnetic and antiferromagnetic interactions where unexpectedly the ferromagnetic  interactions are stronger than the antiferromagnetic ones. This complex Hamiltonian does not correspond to any known spin liquid model and points to new directions in the search for quantum spin liquid behavior.
\end{abstract}


\maketitle

\section{Introduction}

Conventional magnets in condensed matter typically develop long-range magnetic order when cooled to low temperatures \cite{Blu03}. Below their ordering temperature the magnetic moments develop a static component which acts as the order parameter for the phase transition and the spin ordering is observable as magnetic Bragg peaks which characterize the type of order. The excitations are usually spin-waves which are collective oscillations of the spins about their ordering directions. The transition can be described by Landau theory where symmetry is broken  and the new phase is characterized by a local order parameter \cite{Lan37}.

It was recently realized that some states of matter are characterized by topological order, rather than by symmetry breaking and a local order parameter \cite{Kan05}. This important discovery promises new, exotic and potentially useful properties that could be of relevance e.g. to information technologies. For example topological order gives rise to coherent states which can be highly robust to the usual perturbations that destroy coherence in ordinary states due to the non-local nature of their correlations. One potential application is to provide stable qubits in a topological quantum computer \cite{Nay08,Ste13}.

In magnetism topological order is found in the spin liquid state \cite{Wen91}. This state is characterized by the absence of static long-range magnetic order or spin freezing. Instead the spins remain dynamic, highly entangled and in coherent motion down to the lowest temperatures \cite{Bal10}. The excitations of a spin liquid are spinons which are particles with fractional quantum numbers that cannot be created individually but must be produced in numbers greater than one. Spinons are well-understood in one-dimensional (1D) antiferromagnets \cite{Fad81} where they have a spin quantum number of $S=\nicefrac{1}{2}$ which makes them individually inaccessible to condensed matter probes such as neutron or Raman scattering whose selection rules ensure that the spin quantum number changes by an integer amount. As a result spinons have to be created in multiple pairs and are observed as a continuum of excitations in complete contrast to the single spin-wave excitations observed in conventional magnets by inelastic neutron scattering as a sharp mode with a well-defined trajectory throughout energy and wavevector space \cite{Lak05,Lak13,Bal14}. 

The main ingredient for spin liquid behavior is competition which leads to many quasi-degenerate ground states so the system is unable to choose a single configuration but oscillates between the different possibilities. Competition can be achieved when the interactions between the magnetic ions are incompatible with each other and/or with anisotropies present in the system. While these ideas are being explored theoretically using model Hamiltonians, real spin liquid materials are very rare because even small additional terms in the Hamiltonian lift the degeneracy allowing the system to revert to static long-range magnetic order and spin-wave excitations.

The most famous example of a spin liquid that arises from competition between interactions and anisotropies is spin ice as realized in Ho$_2$Ti$_2$O$_7$ and Dy$_2$Ti$_2$O$_7$ where the magnetic ions form a pyrochlore lattice of corner-sharing tetrahedra \cite{Fen09,Mor09}. Here ferromagnetic interactions which favor parallel spin alignment compete with local anisotropies that favor a non-collinear arrangement. The resulting ground state has no static magnetic order although it has a topological 2-in-2-out order on each tetrahedron. The excitations are free monopoles which can be regarded as fractionalized magnetic dipoles. Another example is the Kitaev spin liquid where the magnetic ions form a honeycomb lattice \cite{Kit06}. The highly directional Ising interactions couple just one spin component. For each interaction that a spin makes with its three nearest neighbors a different component is favored leading to the competition that gives rise to the spin liquid ground state and Majorana fermion excitations. Real Kiteav spin liquid candidates such as RuCl$_{3}$ and Na$_2$IrO$_3$ contain additional terms in their Hamiltonians including Isotropic (Heisenberg) interactions and higher neighbor interactions which induce static magnetic order in the ground state \cite{Ban16,Sin12}. Nevertheless signatures of Majorana Fermion excitations are visible at high energies above the ordering temperature \cite{Ban16}. 

Spin liquids based on Heisenberg interactions with no anisotropy rely entirely on competition between the interactions to generate frustration. It is usually assumed that predominantly antiferromagnetic interactions are necessary. This is because spins coupled by ferromagnetic Heisenberg interactions can rotate and align parallel to each other on any lattice, satisfying all the bonds simultaneously and giving rise to ferromagnetic long-range order. Specific lattices are required that allow geometrical frustration such as those consisting of triangular units. Here antiferromagnetic coupling favors antiparallel spin alignment between nearest-neighbor spins, which can never be satisfied on all magnetic bonds. This typically leads to highly degenerate ground states and the tendency for static long-range order is reduced. This tendency can be further suppressed in quantum systems where the magnetic ions have quantum spin number $S=\frac{1}{2}$; here the Heisenberg uncertainty principle produces zero-point motion that is comparable to the size of the spin and which persists down to $T =0$~K \cite{Bal10}.

The most celebrated isotropic quantum spin liquid model is the kagome where $S=\frac{1}{2}$ magnetic ions on a two-dimensional lattice of corner sharing triangles interact via antiferromagnetic nearest neighbor interactions. Among the proposed physical realizations of the kagome, the best candidate is Herbertsmithite, which has recently been verified as a quantum spin liquid \cite{Han12}. Herbertsmithite has no long-range order down to the lowest temperatures, and its excitations are spinons which are observed as a diffuse multi-spinon continuum in inelastic neutron scattering. $S=\frac{1}{2}$ kagome magnets with higher neighbor interactions have also been explored. In particular Kapellasite has a spin liquid ground state. It consists of ferromagnetic first neighbor interactions which compete with antiferromagnetic third neighbor interactions (across the diagonal of the hexagon) that are both isotropic and of approximately equal strength \cite{Fak12}. This system can be viewed as a network of spin-1/2 antiferromagnetic chains coupled together by frustrated interactions \cite{Gon16}.

Recently the new quantum spin liquid compound \cacro was discovered \cite{Bal16}. Using a combination of bulk properties measurements and muon spectroscopy it was shown that the ground state has no static long-range magnetic order or any static magnetism down to $T=19$~mK and that the spins are fluctuating at the lowest temperatures. The spin fluctuations slow down as temperature is reduced and become constant below $T=0.4$~K where \cacro appears to enter a state of persistent spin dynamics. Inelastic neutron scattering revealed broad diffuse features typical of a multi-spinon continuum rather than sharp spin-waves. Further evidence for the spin liquid state was provided by pseudofermion functional renormalization group (PFFRG) calculations which can compute the static susceptibility and determine whether a specific Hamiltonian develops static long-range magnetic order. PFFRG calculations using the Hamiltonian of \cacro clearly showed the absence of static magnetism and furthermore were able to reproduce the diffuse hexagonal ring-like scattering observed in the data as a function of wavevector. Interestingly this spin liquid state was shown to be highly robust and to survive variations in the individual exchange parameters by up to 50\% \cite{Bal16}.

\begin{figure*}
\includegraphics[width=\textwidth]{structure_2cells.pdf}
\caption{\textbf{The structure of \cacro and the possible 2D magnetic models.} (a) The crystal structure of \cacro showing only the positions of the magnetic \Cr ions. The red, purple, light and dark green, and light and dark blue lines depict the 7 nearest-neighbor \Cr-\Cr bonds and the corresponding exchange interactions are labeled in the same colors. The dashed blue lines show the boundary of the unit cell. (b) The kagome single-layer model depicted in the $ab-$plane corresponding to the solid box in (a), (c) the kagome bilayer model corresponding to the dashed box in (a), and (d) the coupled hexagon model corresponding to the dotted box in (a). (e) A snapshot of the possible ground-state spin arrangement of one bi-triangle.\label{fig:struct}}
\end{figure*}

In Ref. \cite{Bal16}, the derivation of the Hamiltonian of \cacro was not given, and one of the goals of this paper is to report how the magnetic interactions were determined. In Part 1 the magnetic Hamiltonian is derived by measuring the excitations under an applied magnetic field strong enough to ensure a fully ordered ferromagnetic state. At this field the excitations transform into spin-waves and can be fitted by linear spin-wave theory to extract the exchange interactions. We show that the $S=\frac{1}{2}$ moments in \cacro are coupled into a kagome bilayer structure by isotropic interactions. A complex combination of several ferromagnetic and antiferromagnetic interactions is found where surprisingly the ferromagnetic interactions are considerably stronger than the antiferromagnetic ones. In part 2 of the paper we explore the phase diagram of \cacro using heat capacity and inelastic neutron scattering measurements for several applied magnetic fields up to the saturation field. The heat capacity reveals no phase transitions, but a number of crossovers are evident which are linked to changes in the magnetic excitations as they evolve from diffuse rings to sharp spin-waves. Finally in the discussion section the phase diagram of \cacro as a function of field and temperature showing the boundaries of the spin liquid phase is proposed. We also explain how the Hamiltonian is able to support a quantum spin liquid ground state. The frustration arises from the direct (vertex to vertex) coupling of ferromagnetically coupled triangles to antiferromagnetically coupled triangles which prevents both triangles from achieving their ideal spin configuration. This frustration motif has not been considered before and points to new directions in the search for spin liquid behavior.

\section{Experimental details}

We prepared powder and single crystal samples of \cacro using the method described in Ref. \cite{Bal16_2}. DC susceptibility, magnetization and heat capacity were measured on single crystals over a range of temperatures and applied magnetic fields. The magnetic excitations were explored using inelastic neutron scattering at low temperatures and up to high magnetic fields on both powder and single crystal samples. All experimental details are given in Appendix A.

\section{Part 1 - The Hamiltonian}

In this section the magnetic Hamiltonian of \cacro is derived by fitting its magnetic excitation spectrum measured by inelastic neutron scattering under a large applied magnetic field to linear spin-wave theory. DC susceptibility, magnetization and heat capacity data are also presented and are used to constrain the spin-wave fitting. The possible magnetic models are discussed and evaluated.

\subsection{Introduction of the models}

\begin{figure}
\includegraphics[width=0.9\columnwidth]{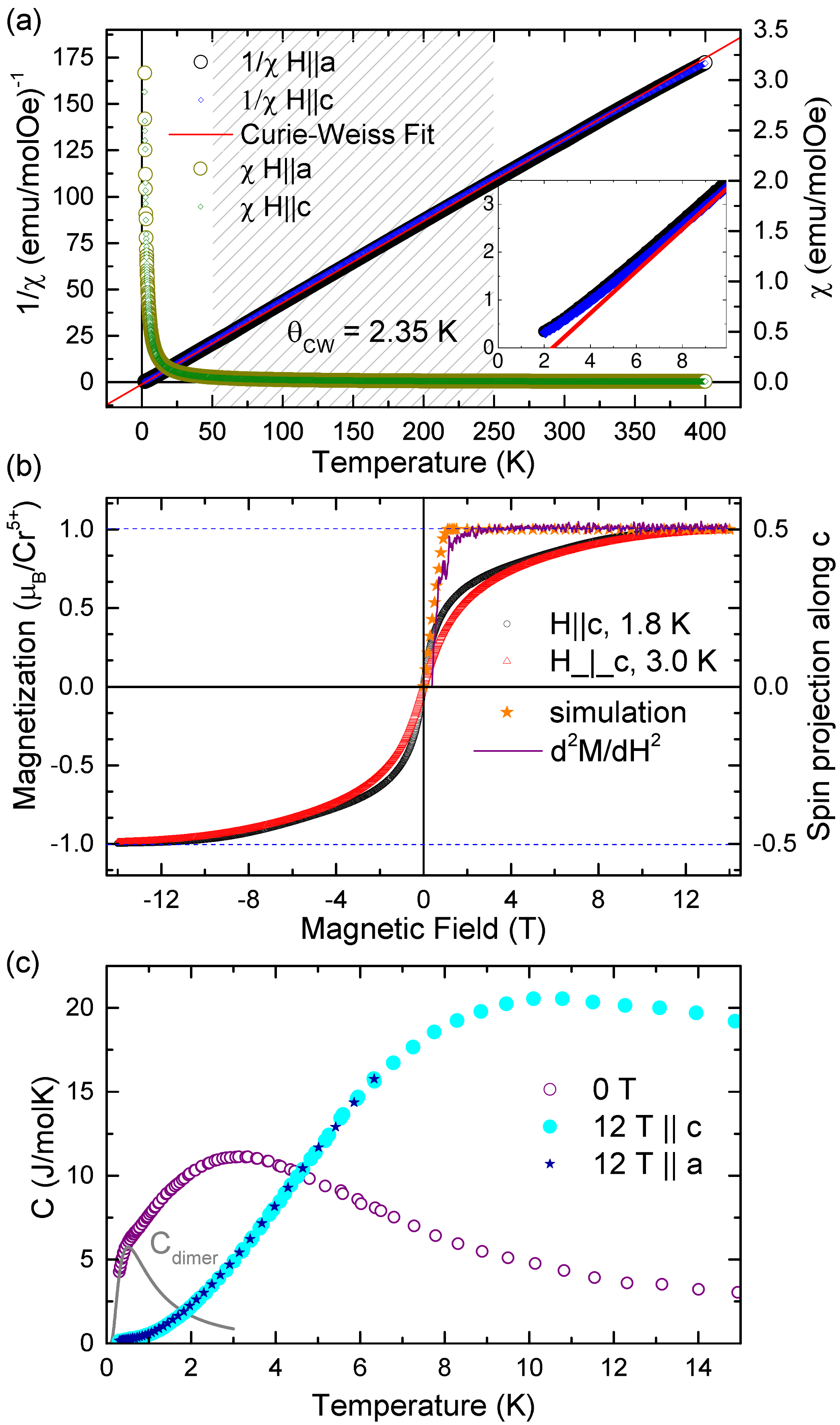}
\caption{\textbf{The bulk properties.} (a) DC susceptibility (green circles) measured with a magnetic field of 0.1~T for field directions $H||a$ and $H||c$, plotted together with the inverse susceptibility (blue and black circles). The red line is a fit to the Curie-Weiss law in the shaded region. The inset shows the behavior at lowest temperatures. (b) Magnetization data measured with $H||c$ at $T=1.8$ K and $H\perp c$ at $T=3$ K. The simulated magnetization obtained from a mean-field calculation at $T=0$~K based on the kagome bilayer model is overplotted.  In addition, the second derivative $\frac{\partial^2M}{\partial H^2}$ for $H||c$ is shown. (c) Magnetic heat capacity for zero field and $H=12$~T (see Appendix B for details of the phonon subtraction). The zero field data below $T=0.46$~K was fitted to the heat capacity of a gapped dimer magnet $C_{\text{dimer}}=R(\Delta/T)^2\text{e}^{(-\Delta/T)}$ (solid gray line) providing an upper limit for the energy gap of $\Delta \leq 0.09$~meV.}\label{fig:bulk}
\end{figure}

The positions of the magnetic Cr$^{5+}$ ions in \cacro are represented in figure~\ref{fig:struct}(a) by the black and gray spheres \cite{Bal16_2}. They form layers in the $ab-$plane that are stacked along the $c-$direction. The seven shortest inequivalent Cr$^{5+}$-Cr$^{5+}$ bonds are represented by the colored lines and the corresponding magnetic exchange interactions are labeled in order of increasing bond distance as J0, J11, J12, J21, J22, J31 and J32 (see table~\ref{tab:Hamiltonian}). The crystal structure can be simplified into several two-dimensional (2D) magnetic models. The consideration of purely 2D models is justified by the absence of magnetic dispersion along the $c$ axis as reveal below by the inelastic neutron scattering data.

Figure~\ref{fig:struct}(b) shows the arrangement of Cr$^{5+}$ ions in a single layer. This model is obtained by setting the interactions J0, J11 and J12 to zero. Here the Cr$^{5+}$ ions form a kagome lattice in the $ab-$plane (2D structure of corner sharing triangles). The kagome layer is distorted because it consists two different equilateral triangles (corresponding to two inequivalent interactions represented by the green and blue bonds) which alternate. Furthermore there are two different kagome layers in \cacro which feature slightly different bond distances indicated by the different hues of blue and green visible in figure~\ref{fig:struct}(a). In one plane the triangles are formed by the J22 (light green) and J31 (dark blue) bonds, while in the other layer they are formed by the J21 (dark green) and J32 (light blue) bonds. 

Figure~\ref{fig:struct}(c) shows the kagome bilayer model obtained by setting J11=J12=0. Here the interaction J0 (red lines) couples together the two inequivalent kagome layers such that the dark blue triangles (J31) in the first layer are coupled directly (vertex to vertex) to the dark green triangles (J21) in the second layer while the light blue triangles (J32) are coupled to the light green triangles (J22). Finally, figure~\ref{fig:struct}(d) shows the coupled hexagon model. It is achieved by setting J0=0 while all other exchange interactions are allowed. The couplings J11 (dark purple) and J12 (light purple) form hexagons which are connected into a triangular lattice by J21 and J22. The interactions J31 and J32 form additional paths within each hexagon.

\subsection{The bulk properties}

\begin{figure*}
\includegraphics[width=\textwidth]{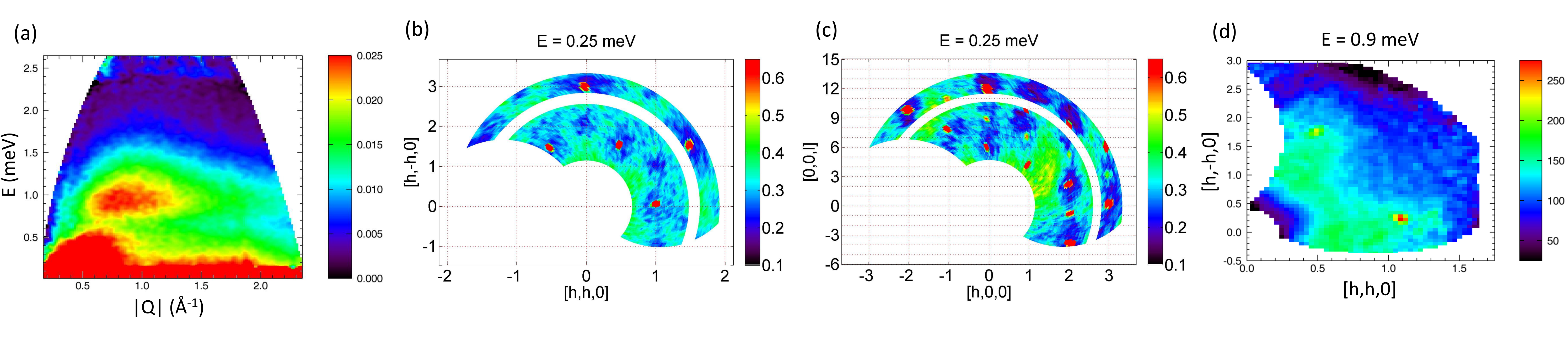}
\includegraphics[width=\textwidth]{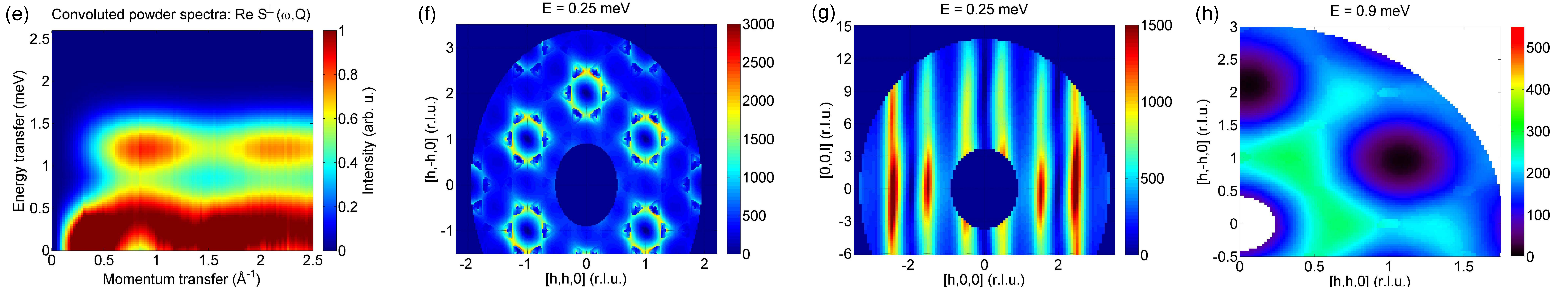}
\caption{\textbf{Powder and single crystal inelastic neutron scattering in zero field together with spin-wave simulations.} (a) Powder data measured on TOFTOF at the temperature $T=430$~mK as a function of energy and wavevector transfer; the high intensity below $E=0.15$~meV is due to coherent and incoherent elastic scattering.  Single crystal data measured on IN14 at $T=1.6$~K for energy transfer $E=0.25$~meV in the (b) $(h,k,0) $plane and (c) $(h,0,l) $plane.  (d) Single crystal data measured on MACS at $T=0.09$~K for $E=0.9$~meV in the $(h,k,0) $plane. The red spots of very high intensity in (b-d) are due to phonons dispersing out of nuclear Bragg peaks. (e-h) Spin-wave simulations corresponding to the datasets (a-d). In the simulations the spectrum was broadened by an energy width of 0.4~meV in (f,g) and 0.5~meV in (h). The constant energy simulations shown in (f) and (g) were centered at 0.25~meV and in (h) at 0.9~meV.}\label{fig:diffuse_scattering}
\end{figure*}

Magnetic and thermodynamic bulk property measurements can give important insights into the spin Hamiltonian. DC susceptibility, magnetization, and heat capacity measurements of \cacro are presented in figure~\ref{fig:bulk}. The DC susceptibility measured between 400~K and 1.8~K (Fig.\ \ref{fig:bulk}(a)) increases smoothly as temperature decreases and there is no indication of a magnetic phase transitions in this temperature range. The smooth increase is also inconsistent with a gap in the excitation spectrum which would be observed as a drop in susceptibility at lowest temperatures. A spin gap can occur due to magnetic anisotropy or because there is a dominant antiferromagnetic interaction which pairs the magnetic ions into spin singlets (dimers) leading to a non-magnetic ground state and gapped excitation. Furthermore the magnetic interactions appear to be isotropic since the susceptibilities for $H||a$ and $H||c$ lie precisely on top of each other.

The inverse susceptibility was fitted to the Curie-Weiss law in the temperature range 50-250~K. The effective moment per \Cr ion is $\mu_{eff} = 1.74(2) \mu_B$ confirming that they have spin $S = \frac{1}{2}$ ($\mu_{eff} = g_s \sqrt(S(S+1)) \mu_B = 1.73 \mu_B$ for $S=\frac{1}{2}$ and $g_s=2$). The Curie-Weiss temperature is only $+2.35$~K ($\approx0.2$~meV) which indicates that the magnetic interactions are either extremely weak and/or that there is a mixture of ferromagnetic (FM) and antiferromagnetic (AFM) interactions which partly cancel each other. The presence of AFM interactions is also evidenced by the slight upturn of the inverse susceptibility at lowest temperatures.

The magnetization increases rapidly at low fields, then more gradually at larger fields, finally achieving saturation at around $H_{sat}=12$~T (Fig.\ \ref{fig:bulk}(b)). The saturation field suggests that \cacro has a magnetic energy scale of $\sim g_s \mu_B S H_{sat} = 1.4$~meV (assuming $g_s = 2$). The shape of the magnetization curve is consistent with a mixture of FM and AFM interactions. The difference between the two datasets can be ascribed to the different measurement temperatures. No features indicating a field-induced transition are visible. The magnetization also does not suggest a significant spin gap which would suppress the magnetization below a critical field, however it should be noted that the temperature of the measurement was relatively high.

The magnetic heat capacity measured in zero field shows no transition to long-range magnetic order down to 300~mK (Fig.\ \ref{fig:bulk}(c)). However a broad maximum centered at $\sim$3~K indicates the onset of short range magnetic correlations. The heat capacity shows no indication of a spin gap which would be observed as a suppression of the heat capacity at lowest temperatures and an exponential increase with temperature. Fitting the region below 0.46~K to the heat capacity of a gapped dimer magnet (see caption of Fig.\ \ref{fig:bulk}) yields a maximum possible gap of 0.09~meV. The heat capacity in a magnetic field of 12~T is also shown for field applied parallel to the $c$ and $a$ axes. No difference is observed between these two field directions again suggesting isotropic, Heisenberg interactions. At this field which is the field where the magnetization saturates (Fig.\ \ref{fig:bulk}(b)), the broad maximum observed at $\sim$3~K in zero field has disappeared and the heat capacity drops smoothly to 0 as T$\rightarrow$0~K. This is consistent with the saturated state where the spins are all parallel to the field and no magnetic entropy is left around the ground state. A broad maximum centered at 10.5~K remains which is caused by a Schottky-like contribution due to the excitations that are gapped by the external field.

\subsection{Zero-field inelastic neutron scattering}

Inelastic neutron scattering data provides further information about the magnetic Hamiltonian. Powder inelastic neutron scattering in zero-field shows two magnetic excitation bands extending over the energy ranges 0-0.6 meV and 0.7-1.6 meV (Fig.~\ref{fig:diffuse_scattering}(a)). The excitations appear gapless within the instrumental resolution and are broad at all wavevectors in contrast to spin-waves which would have a well-defined dispersion at lowest wavevectors in a powder measurement. The upper energy boundary of the magnetic spectrum is approximately $E_{max}=1.6$~meV. This can be used to calculate the saturation field assuming $E_{max}=g_s \mu_B S H_{sat}$ to give $H_{sat}=13.8$~T, which is close to the saturation field of $H_{sat}\approx 12$~T observed in the magnetization measurements.

The diffuse character of the excitations is further confirmed by single crystal inelastic neutron scattering measurements. Broad excitations with 6-fold symmetry are observed in the $(h,k,0)$ scattering plane (kagome plane) at all energies and wavevectors. Figures~\ref{fig:diffuse_scattering}(b) and (d) show typical slices through the excitations in the lower (0.25 meV) and upper (0.9 meV) bands respectively, further data is shown in Ref.~\cite{Bal16}. Spin-wave excitations which are the excitations of conventional magnets and would be observed as sharp rings can be ruled out. In Ref.~\cite{Bal16} the magnetic scattering of \cacro was ascribed to spinon excitations. Spinons possess $S=\nicefrac{1}{2}$ and cannot be created singly but must be excited in multiple pairs in a neutron scattering process; as a result a broad multi-spinon continuum is observed rather than the single spinon dispersion. Spinons are believed to be the fundamental excitations of a quantum spin liquid. 

Finally it should be noted that the excitations are non-dispersive perpendicular to the kagome plane along the $(0,0,l)$ direction (Fig.~\ref{fig:diffuse_scattering}(c)). This reveals that the magnetic correlations are confined to the $ab-$plane and justifies the consideration of only 2D magnetic models in figure~\ref{fig:struct}. 

\subsection{Inelastic neutron scattering at high magnetic field}

\begin{figure}
\includegraphics[width=0.9\columnwidth]{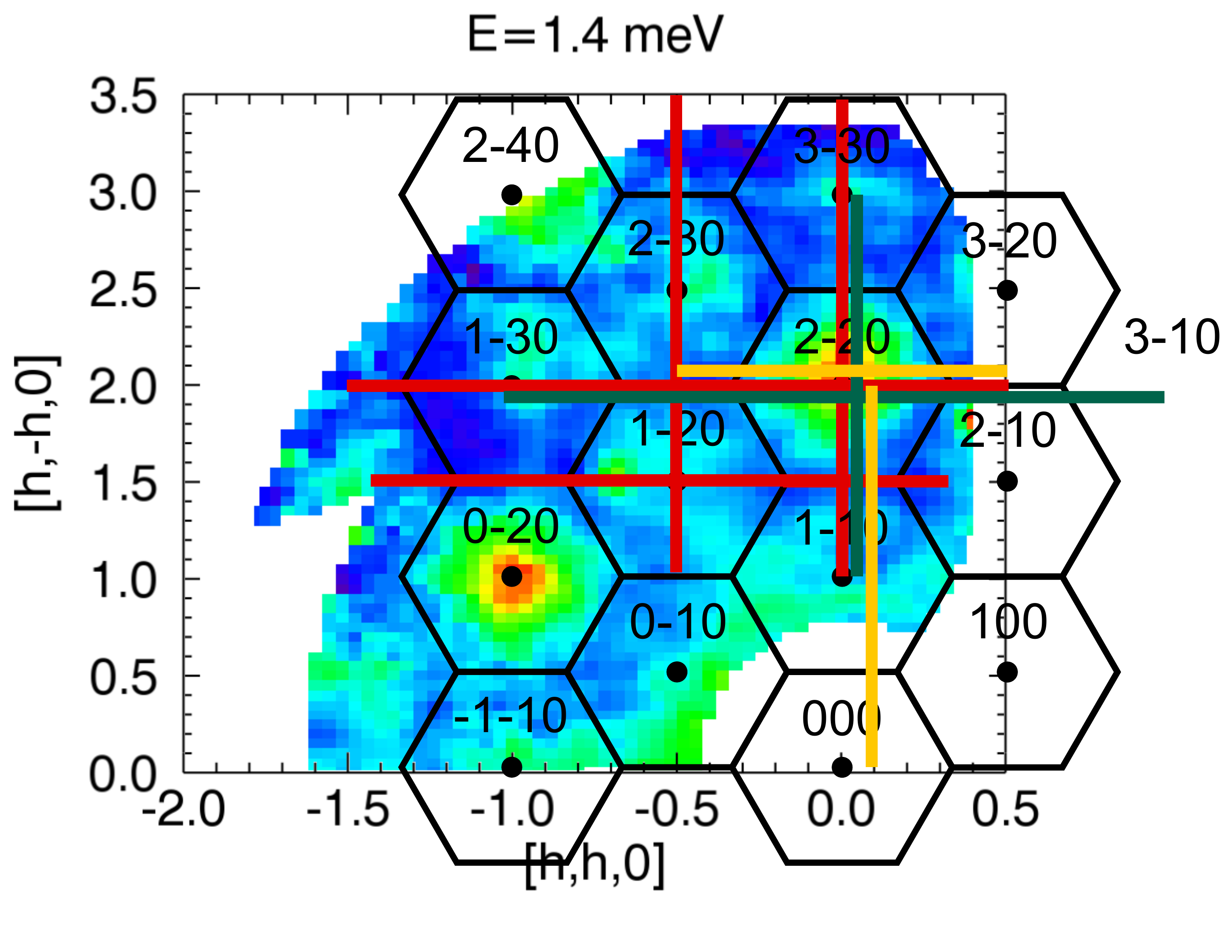}
\caption{\textbf{The scattering plane of \cacro.} A constant energy slice in the $(h,k,0)$ scattering plane centered at $E=1.4$~meV measured on the MACS spectrometer at $H=11$~T and $T=0.09$~K. The reciprocal lattice showing the hexagonal boundaries of the first Brillouin zone (black lines) and the zone centers (black dots) is overplotted. The green, red and yellow lines indicate the directions and wavevector ranges of the $E$ vs.\ $\bm{Q}$ slices measured on the LET, MACS and OSIRIS spectrometers which are presented in figures \ref{fig:spin_waves_9T}, \ref{fig:spin_waves_11T} and \ref{fig:osiris} respectively.\label{fig:scatt_plane}}
\end{figure}

\begin{figure*}
\includegraphics[width=0.75\textwidth]{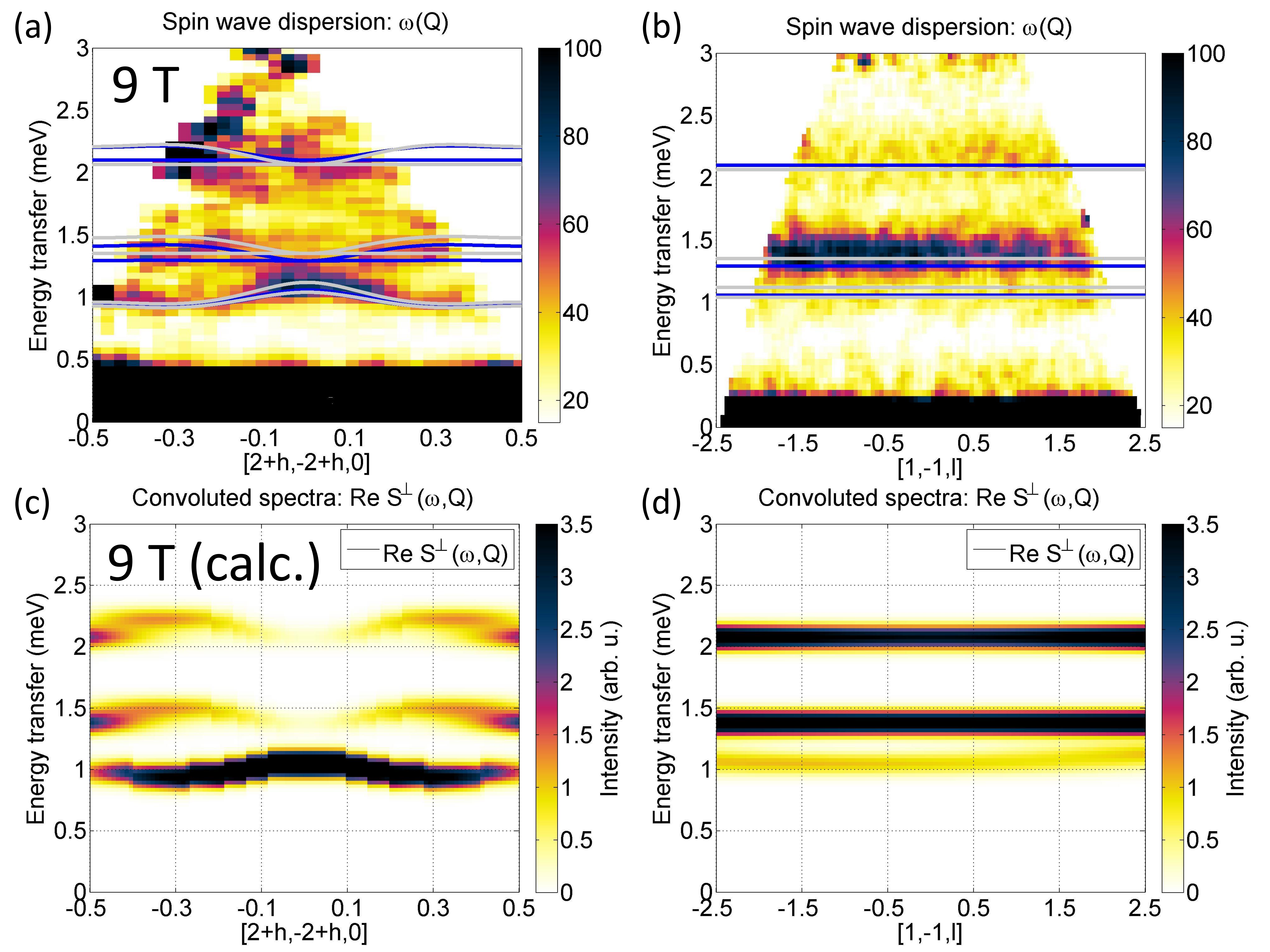}
\caption{\textbf{High field inelastic neutron scattering data compared to spin-wave simulations for the kagome bilayers model.} The upper panels give the $S(\bf{Q},\omega)$ measured on LET at $T=2$~K and $H=9$~T as a function of energy and wavevector transfer along (a) the in-plane $[2+h,-2+h,0]$-direction and (b) the out-of-plane $[1,-1,l]$-direction. The high intensity near $E=0$ is due to coherent and incoherent elastic scattering. The spin-wave dispersions simulated by SpinW at 9~T using the fitted Hamiltonian for the kagome bilayers model are over-plotted. The lower panels show the simulated $S(\bf{Q},\omega)$ along (c) $[2+h,-2+h,0]$ and (d) $[1,-1,l]$ convolved with the Gaussian energy resolution of FWHM = 0.3~meV.}\label{fig:spin_waves_9T}
\end{figure*}

\begin{figure*}
\includegraphics[width=0.8\textwidth]{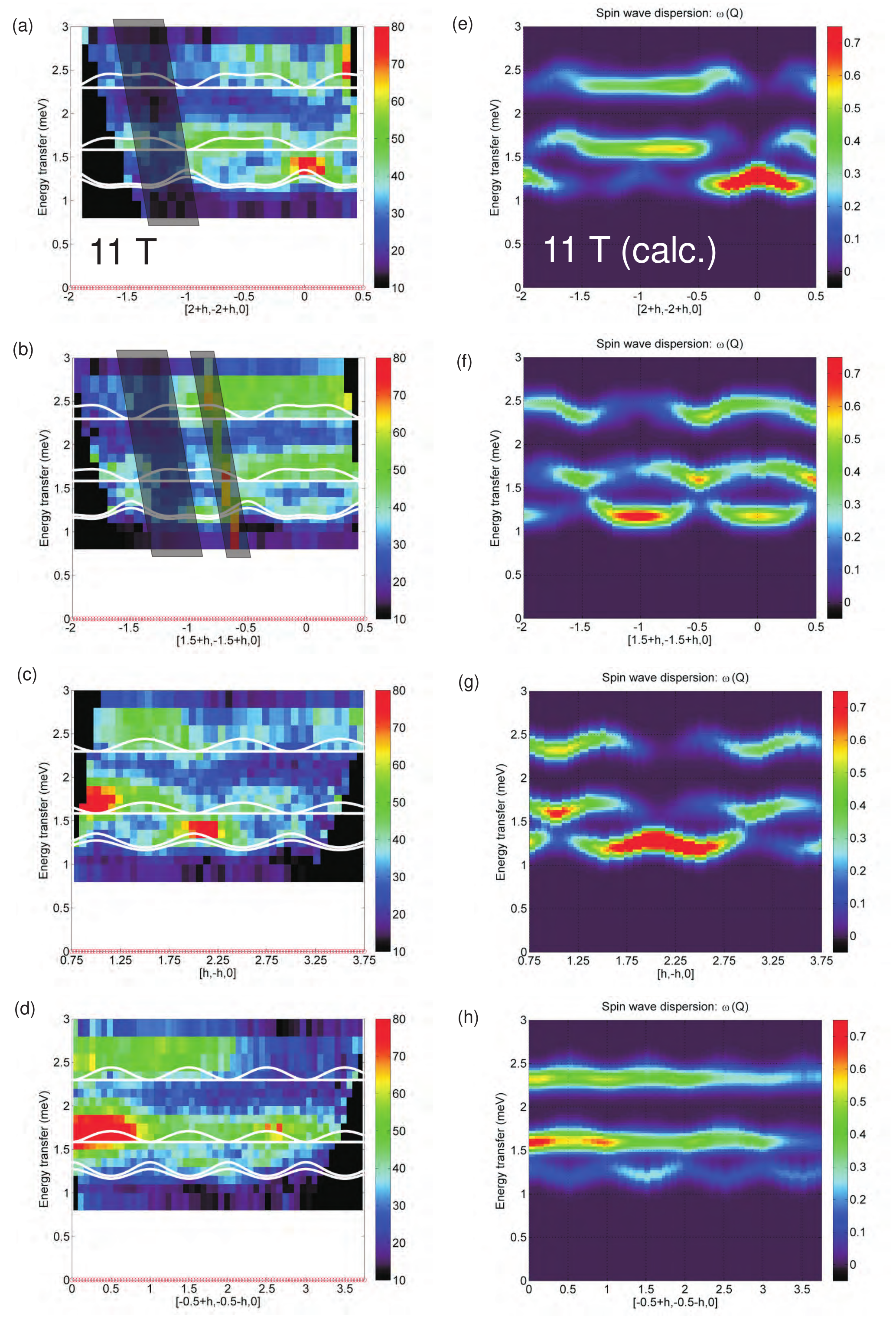}
\caption{\textbf{High field inelastic neutron scattering along with spin-wave fits for the kagome bilayer model.} The panels on the left hand side give the $S(\bf{Q},\omega)$ measured on MACS at $T = 90$~mK and $H = 11$~T as a function of energy and wavevector transfer along various directions in the $(h,k,0)$ plane. The data in the range $0.8<E<2.4$ was collected with an energy step of 0.1~meV, while the energy step of the higher energy data was 0.2~meV. The large shaded regions in (a) and (b) have artificially low intensity due to attenuation by the supporting structures in the magnet. The small shaded region in (b) has higher intensity due a phonon dispersing out of a nuclear Bragg peak. The spin-wave dispersions simulated by SpinW at 11T using the fitted Hamiltonian for the kagome bilayer model are overplotted. The panels on the right-hand side show the calculated $S(\bf{Q},\omega)$ along the same directions convolved with the Gaussian energy resolution of FWHM = 0.4~meV.\label{fig:spin_waves_11T}}
\end{figure*}

Although \cacro does not have long-range magnetic order or spin-wave excitations in zero magnetic field, an applied field equal to the saturation field will force all the spins to align in the field direction and transform the excitations into spin-waves about this ferromagnetic ground state. Figure~\ref{fig:scatt_plane} shows a constant energy slice at 1.4~meV measured at 11~T (close to the experimentally observed saturation field of $\approx$12~T). The contrast with the zero field scattering is immediately evident. The spectrum which was  formerly broad and diffuse at all energies and wavevector transfers is now concentrated mostly in highly intense spots at well-defined reciprocal lattice positions. These correspond to the spin-wave dispersions cutting through the constant energy plane. 

The existence of spin-wave excitations becomes even clearer in energy versus wavevector transfer slices. Data measured at 9~T and at 11~T are shown in figures \ref{fig:spin_waves_9T} and \ref{fig:spin_waves_11T} respectively. The reciprocal space directions in which these slices are taken are indicated by the colored lines in figure~\ref{fig:scatt_plane}. At least three modes are evident which are all gapped by the Zeeman energy of the external field. 
Since the energy resolution was 0.3~meV and 0.35-0.65~meV for the measurements in figures \ref{fig:spin_waves_9T} and \ref{fig:spin_waves_11T} respectively, these excitations correspond to resolution limited spin-waves. All three modes show a clear dispersion in the $(h,k,0) $plane. The spectrum reveals a characteristic `knot'\--like feature at (2,-2,0) and equivalent positions which is centered at $E=1.2$~meV at 9~T and $E=1.4$~meV at 11~T. This is a point of high scattering intensity where the two lowest excitation branches meet. No dispersion is visible perpendicular to the $(h,k,0) $plane along the $[0,0,l]$direction (Fig.~\ref{fig:spin_waves_9T}(b)), indicating that there are no continuous connected magnetic exchange paths along the $c$ direction in \cacro. This again confirms the 2D nature of the magnetic coupling in the $(h,k,0)$ plane.

\subsection{Linear spin wave theory at high field}

Since resolution limited spin wave modes have been observed in the high-field phase, linear spin wave theory can be fitted to the data to extract the exchange interactions. For this purpose the Matlab SpinW library \cite{Tot15} was used and the technical details of the spin wave fitting procedure are described in Appendix C. The following constraints deduced from the bulk properties measurements, were used to obtain the Hamiltonian: 1) The interactions are isotropic (Heisenberg); 2) Both FM and AFM interactions are possible; 3) No individual interaction has a strength greater than 1.4~meV; 4) The resulting magnetic model is 2D in the $ab$-plane. Further, we restricted ourselves to the seven nearest-neighbor exchange paths shown in Fig.~\ref{fig:struct}(a). 

The Heisenberg Hamiltonian under external magnetic field $H$ applied along the $c$-axis takes the form
\begin{align}\label{eq:Ham}
\mathcal{H}=\sum_{i>j}J_{ij}\bm{S}_i\bm{S}_j + \sum_i g_s \mu_B H S_i^c.
\end{align}
where $g_s$ is the g-factor and $\mu_B$ is the Bohr magneton. To determine the exchange parameters $J_{ij}$, the theoretical structure factor $S(\bf{Q},\omega)$, calculated via linear spin wave theory using the Hamiltonian \eqref{eq:Ham} was fitted to the inelastic neutron scattering data. The dispersions were extracted from the data and the theoretical structure factor was compared to them using a non-linear least squares minimization routine. The structure factor was calculated for each of the three magnetic models shown in figure~\ref{fig:struct}: a) the kagome single layer (J0=J11=J12=0), b) the kagome bilayer (J11=J12=0) and c) the coupled hexagons (J0=0). The kagome single layer model was not able to provide an acceptable description of the data, however both the kagome bilayer and the coupled hexagon models achieved good fits of similar quality. The exchange couplings from the best fits for both these models are given in Table~\ref{tab:Hamiltonian}. Here, uncertainties in the Hamiltonian were determined by visual comparison of the theoretical and experimental inelastic structure factors.

\begin{table}
\caption{The exchange couplings of the seven shortest Cr$^{5+}$-Cr$^{5+}$ bonds for the best fits of the kagome bilayer and coupled hexagons models. The goodnesses of fit were $R_w=0.3355$ and $R_w=0.3359$ respectively. Negative interactions are FM while positive interactions are AFM. \label{tab:Hamiltonian}}
\begin{ruledtabular}
\begin{tabular}{l l l l l l}
bond dist- & \multicolumn{2}{l}{kagome bilayer model} & \multicolumn{2}{l}{coupled hexagon model}\\
ance [\AA] & Parameter & Value [meV] & Parameter & Value [meV]\\
\hline
3.883 & $J0$  & -0.08(4) & J0  & 0\\
4.103 & $J11$ & 0 & J11 & 0.02\\
4.167 & $J12$ & 0 & J12 & -0.04\\
5.033 & $J21$ & -0.76(5) & J21 & -0.78\\
5.095 & $J22$ & -0.27(3) & J22 & 0.09\\
5.679 & $J31$ & 0.09(2) & J31 & -0.25\\
5.724 & $J32$ & 0.11(3) & J32 & 0.08\\
\hline
& $\sum{J}$ & -0.91(17) & $\sum{J}$ & -0.88\\ 
\end{tabular}
\end{ruledtabular}
\end{table}

The resulting spin-wave dispersion and structure factor calculated for the kagome bilayer model from the parameters in Table~\ref{tab:Hamiltonian} are compared to the 9~T and 11~T data in figures~\ref{fig:spin_waves_9T} and \ref{fig:spin_waves_11T} respectively. The overall intensity of the spin wave modes was scaled to agree with the measured intensity at the position of the `knot'\ at (2,-2,0). The fitted parameters correctly reproduce the energy of the spin-wave dispersion while the relative intensities are also in agreement except for a few mismatches. The good agreement between our measurement and the linear spin-wave calculation confirms the accuracy of this magnetic model and shows that the high-field state can be described semiclassically. The coupled hexagons model calculated with the parameters listed in Table~\ref{tab:Hamiltonian} also reproduces the inelastic neutron scattering data to high accuracy as shown in Appendix C.

From this analysis it is not possible to determine whether \cacro realizes the kagome bilayer or coupled hexagons model, both give an equally good account of the data. A closer inspection of these two models reveals that despite their superficial differences they in fact have several striking similarities. Both consist of triangles formed by strong ferromagnetic interactions ($\approx$~-0.77 and $\approx$~-0.26) which are coupled directly (vertex to vertex) to triangles formed by weaker antiferromagnetic interactions ($\approx$~0.10 and $\approx$~0.085). In the case of the kagome bilayer model this direct coupling is achieved by the intrabilayer interaction J0=-0.08, while in the coupled hexagons it is achieved by the intrabilayer interactions J11=0.02 and J12=-0.04. Thus both models have similar frustration motifs and interaction strengths. 

Based on the crystal structure, the kagome bilayer model is much more physically reasonable than the coupled hexagons model. This is because the pair of bonds J21 and J22 which have a similar Cr$^{5+}$-Cr$^{5+}$ distances and electronic environments, have similar strengths and the same sign (both ferromagnetic) in the kagome bilayer model whereas they have opposite signed in the coupled hexagons model (see Fig.~\ref{fig:struct} and Table~\ref{tab:Hamiltonian}). The same is true for the pair of bonds J31 and J32. Hence it is concluded that \cacro realizes the kagome bilayers with the interactions listed in Table~\ref{tab:Hamiltonian}.

\section{Part 2 - Behavior at intermediate fields}

In this section we explore how the quantum spin liquid ground state and diffuse excitations of \cacro that occur under zero applied magnetic field evolve into the ferromagnetically order state and spin-wave excitations observed at high magnetic field.

\subsection{Heat Capacity}

\begin{figure}
\includegraphics[width=\columnwidth]{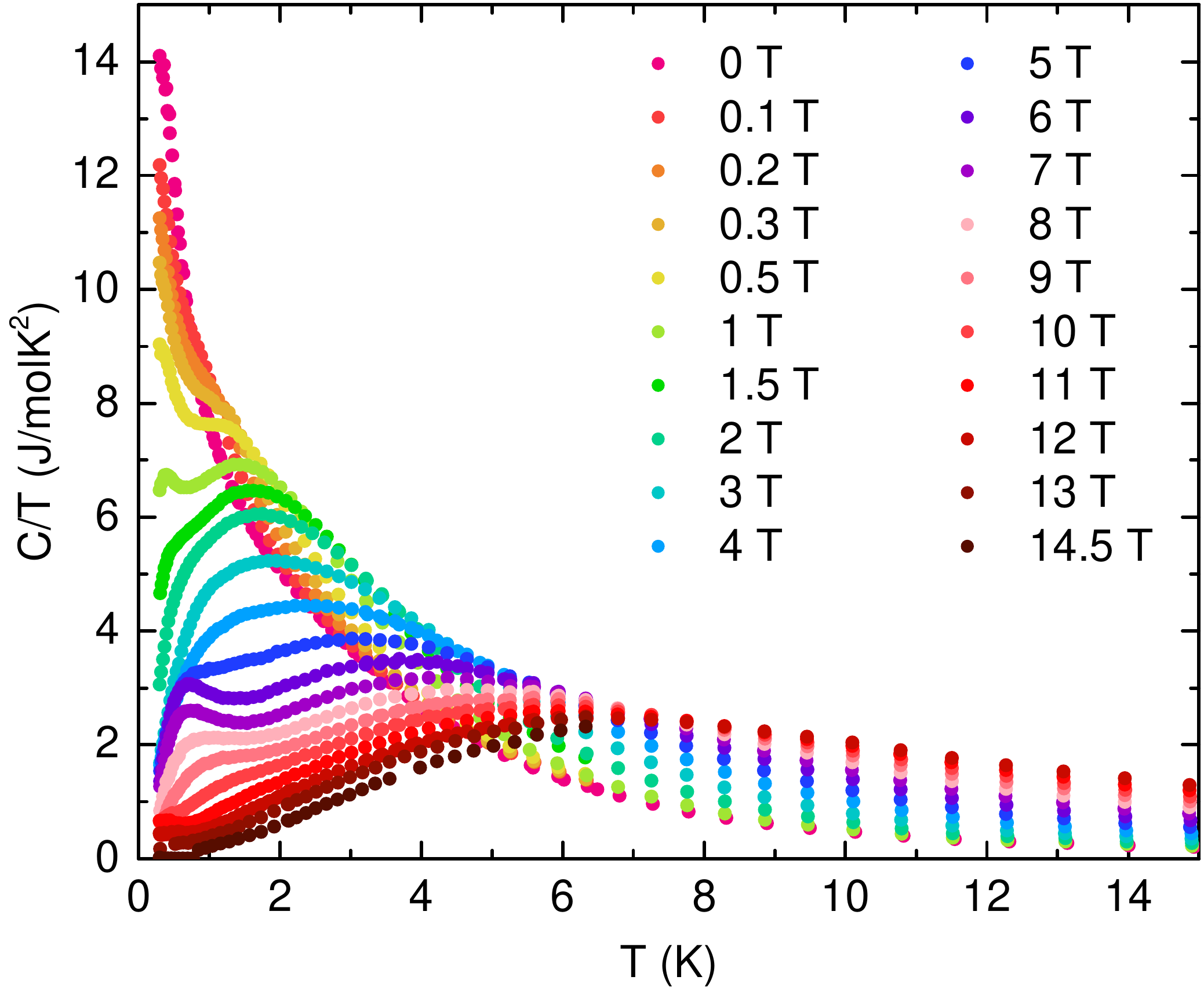}
\caption{\textbf{Magnetic heat capacity divided by temperature ($C_p/T$) as a function of temperature for several applied magnetic fields.} The data was collected for temperatures down to $T=0.3$~K and magnetic fields up to $H=14.5$~T applied parallel to the $c$ axis (almost identical results were achieved for $H$ parallel to the $a$-axis). The phonon contribution subtraction is described in Appendix B. \label{fig:cp}}
\end{figure}
 
Figure~\ref{fig:cp}(a) shows the magnetic heat capacity divided by temperature $C_p/T$, plotted as a function of temperature down to $T=0.3$ K for a number of applied magnetic fields up to $H=14.5$~T. No Lambda anomalies are observed, indicating the absence of any magnetic phase transition, nevertheless several interesting features are visible over this field and temperature range. In zero field a smooth and rapid increase in $C_p/T$ is observed with decreasing temperature. This is reduced for small fields and for $H=1$ T, a weak peak is visible at $T=0.4$ K indicating a distinct crossover in the magnetic ground state. Traces of this crossover are also present from $H=0.5$ to $1.5$~T but at higher fields it broadens and disappears. At $H=6$~T a new peak in $C_p/T$ emerges at $T \approx 0.7$~K possibly indicating another crossover, it broadens with increasing field and finally disappears at $H=10$ T.

Above $H=11$~T, $C_p/T$ is flat and close to zero at the lowest temperatures and shows an exponential increase with temperature indicating that the excitations are now cleanly gapped.
A Schottky-like broad peak is visible, this peak occurs at $T \approx 6$~K for $H=11$~T suggesting a energy gap of $\Delta \approx 1.6$ meV between the ground state and lowest excitations for this applied field. The peak position increases linearly with increasing field indicating that the gap widens following a Zeeman-type dependence. It should be noted that this peak is also visible at lower fields e.g.\ at $H=6$~T it appears at $T \approx 4.0$~K, although for fields below $H=11$~T the excitations are not cleanly gapped and additional features are observed in the low temperature $C_p/T$.

\subsection{Inelastic neutron scattering}

\begin{figure*}
\begin{sideways}
\includegraphics[width=1.2\textwidth]{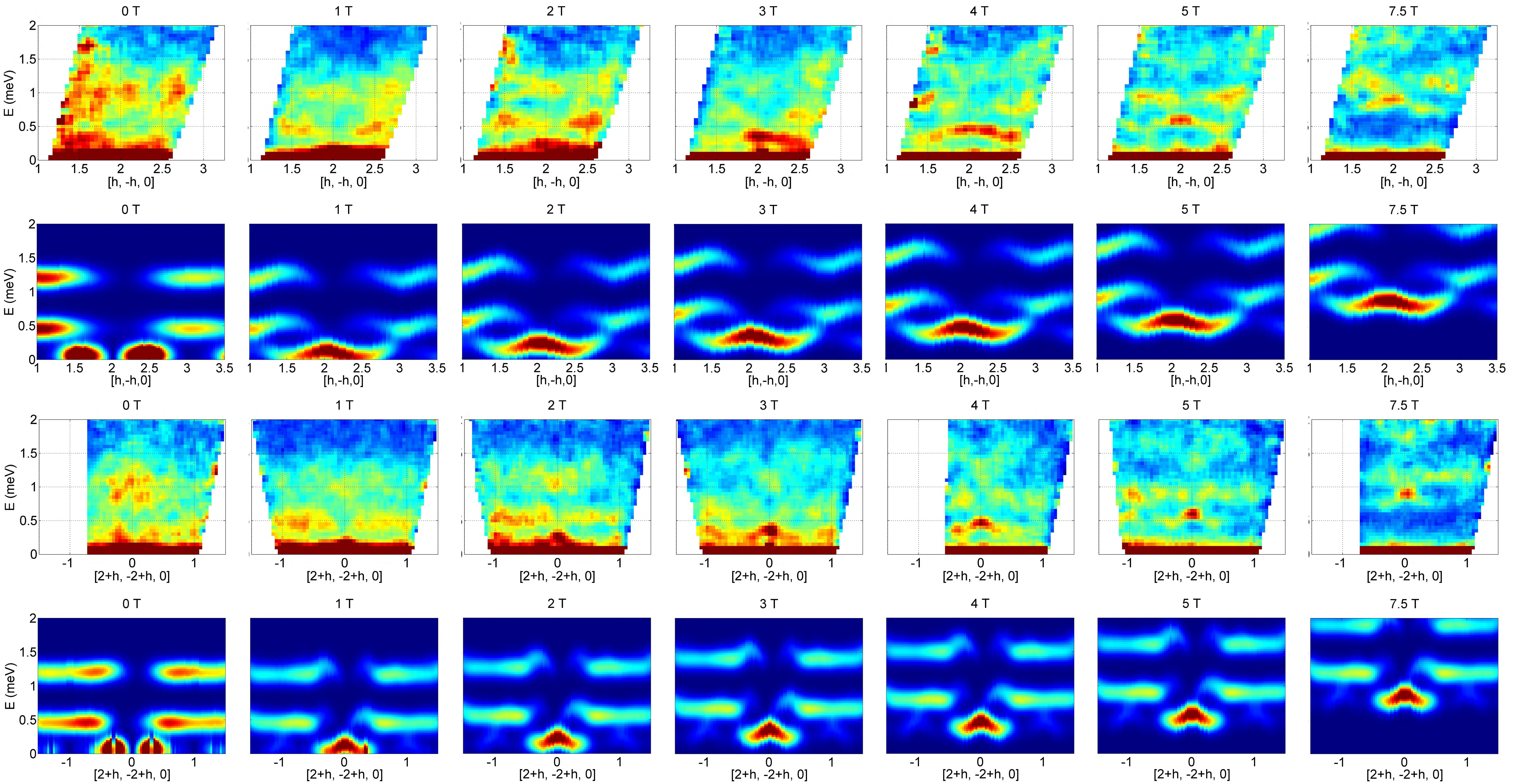}
\end{sideways}
\caption{\textbf{Inelastic neutron scattering and spin-wave simulations under various applied magnetic fields.} (first and third rows) Slices of the inelastic neutron scattering data measured on the OSIRIS spectrometer, plotted as a function of energy transfer versus wavevector transfer along [h,-h,0] (first row) and [2+h,-2+h,0] (third row). The data was collected for temperatures $T\leq220$~mK under the external magnetic fields labeled in the subplots which were applied parallel to the $c$-axis. The strong signal for energies 0-0.2 meV is incoherent scattering. (second and fourth row) Calculated spin wave spectrum for the kagome bilayer model using the Hamiltonian parameters extracted by fitting the 11~T dataset (Table~\ref{tab:Hamiltonian}). Although the instrumental energy resolution is 0.025-0.030~meV, the calculated $S(\bf{Q},\omega)$ is convolved with a Gaussian of FWHM 0.25~meV for a realistic comparison to the data.\label{fig:osiris}}
\end{figure*}

To gain more insight into the different phases of \cacro observed in the heat capacity as a function of magnetic field at low temperatures, single crystal inelastic neutron scattering was measured under several applied magnetic fields. Figure~\ref{fig:osiris} (first and third row) show the excitation spectrum measured in the $(h,k,0)$ plane at $T\leq220$~mK for magnetic fields between $H=0$ and 7.5~T. The evolution from diffuse excitations at zero field to relatively well-defined modes at $H=7.5$~T is evident. For zero field, the intensity is gapless and extends up to an energy of 1.7~meV while the spectral weight is distributed between two broad bands extending over $0<E<0.6$~meV and $0.6<E<1.7$~meV respectively, in agreement with the powder data displayed in Fig.~\ref{fig:diffuse_scattering}(a). The excitations are diffuse at all energies and wavevectors, and no well-defined features are found. 

The scattering changes drastically in an external magnetic field of $H=1$~T. The spectral weight shifts and becomes more well-defined revealing a broad and diffuse mode at an energy of $\sim$0.5~meV. Further diffuse scattering remains at higher energies centered at about 1~meV. An additional feature starts to become visible at the lowest energies emerging from the ground state at the (2,-2,0) position. This is the wavevector of the characteristic `knot' observed at $H=9$ and $11$~T where the two lowest bands touch (see Fig.~\ref{fig:spin_waves_11T}). These three features sharpen and shift to higher energy with increasing field. At $H=2$~T the lowest energy mode becomes more visible and by $H=5$~T it has shifted to an energy of $\sim$0.5~meV and is clearly separated from the elastic scattering. At this field the two higher energy modes now lie at $\sim$1.0~meV and $\sim$1.5~meV respectively, all modes remain broad and lie on a diffuse scattering background which extends up to 1.7~meV. At $H=7.5$~T the upper band has partly moved out of the measured energy range, while the lower two modes start to develop the characteristic features observed at $H=9$ and $11$~T such as the `knot' at (2,-2,0). Furthermore the diffuse scattering below the lowest energy mode is considerably reduced so that an energy gap of 0.6~meV is visible. Even at this field the modes are much broader than the instrumental resolution showing that they cannot be described as pure spin waves.

The development of the magnetic excitations under external field can be compared to the field dependence of the heat capacity. The distinct weak peak in $C_p/T$ at $H=1$~T and $T=0.4$~K (Fig.~\ref{fig:cp}) appears at the same field at which the lowest energy mode starts to emerge from the ground state, indicating a crossover from the zero field spin liquid into a different phase. In addition the emergence of the Schottky-type broad peak can be linked to the opening of the energy gap in the excitation spectrum that is already partially formed at $H=7.5$~T. Finally the broad peak that is observed from $H=6$ to $10$~T at $T \approx 0.7$~K suggests the presence of low energy states below the energy gap at an energy of $\approx 0.06$~meV. While such states are not clearly evident in the magnetic excitations spectrum they may be masked by the strong incoherent scattering that extends up to 0.2 meV. Alternatively these states may be invisible to neutron scattering e.g.\ they might be singlet states.

\subsection{Simulations at intermediate fields}

In order to deepen our understanding of \cacro, we compared it to a mean-field ground state and spin-wave excitations. A ground state with long-range magnetic order is of course not valid at these low fields and as already pointed out, the observed diffuse excitations cannot be described as spin-waves which are well-defined at all energies and wavevectors. We therefore expect that these calculation can be used only as a guide to the approximate spin configuration and the general energy scale and intensity of the excitations. In order to find an effective ground state we used the following method: first the external magnetic field was added to the Hamiltonian for the kagome bilayers model and the magnetic structure was optimized by minimizing its energy assuming a single ordering wavevector. At $H=14$~T this leads to a ferromagnetic structure with the spins entirely parallel to the external field. We then gradually decreased the field in the simulations and optimized the magnetic structure at each field step using the structure from the previous step as the starting point for the optimization. A step size of $dH=0.5$~T was used for fields greater than $H=1.5$~T, while a step size of $dH=0.1$~T was used at lower fields.   

To make comparison to the magnetization data, we extracted the average spin projection long the external field direction of the mean-field magnetic structure as a function of field. This quantity is plotted over the magnetization data in Fig.~\ref{fig:bulk}. The calculation suggests a fully polarized phase for fields $H > 1$ in contrast to the data which smoothly decreases below $H=11$~T. Below $H=1$~T the calculated magnetization decreases steeply and linearly to zero. The measurement also shows a change of slope at $H=1$~T albeit a much more gradual one (see also the second derivative $\frac{\partial^2M}{\partial H^2}$ for $H||c$ in Fig.~\ref{fig:bulk}(b)) and the data and simulation match each other at lowest fields. The clear discrepancies between the simulation and the data may be caused by the high measurement temperatures of $T=1.8$~K and $T=3.0$~K, in contrast the calculation was performed at $T = 0$ K. 
The elevated temperatures would smooth sharp features in the magnetization as is already clear from a comparison of the data measured at $T=1.8$~K and $T=3.0$~K. Magnetization measurements at dilution temperatures may provide better agreement between data and theory. Nevertheless it is interesting to see that the change of slope observed in both the data and the theory occurs at $H=1$~T where $C_p/T$ shows a peak and the inelastic neutron scattering data also reveal the formation of a mode at lowest energies, again suggesting that a crossover occurs at this field.

Spin-wave theory based on this mean-field ground state was used to calculate the excitations energies and neutron scattering cross-sections for \cacro at various fields. In figure~\ref{fig:osiris} the spin-wave calculations are plotted alongside the data in the second and fourth rows. Well-defined spin-wave modes are predicted for all fields in striking contrast to the diffuse features observed. It should be mentioned that to make comparison with the data the spin-waves were broadened by a Gaussian with FWHM=0.25~meV, while the instrumental energy resolution is only 0.025-0.030 meV. Thus significant quantum fluctuations must be still present at these intermediate fields which broaden the spectrum. Nevertheless as the field is increased and the excitations in the data sharpen, the agreement between the spin-wave simulations and the measurement continuously improves. The energies and intensities of the simulated spin-waves actually correspond rather well to the energies where the scattering is strongest in the data and capture how the modes develop and shift upwards in energy with increasing field. In zero field the lowest energy mode is gapless in the simulation in general agreement with the rather intense scattering observed down to lowest energies in the data. This mode develops a dispersion at $H=1$~T in agreement with the data and becomes gapped for fields $H > 1$~T. The clean gap predicted by the simulations is not observed in the data for fields up to $H=5$~T. At $H=7.5$~T the gap becomes cleaner while the excitations become clearer and in particular the `knot' observed at (2,-2,0) is nicely reproduced by the spin-wave simulation. 

\subsection{Simulations at zero field}

As expected, the agreement between the spin-wave calculation and the data is worst in zero-field. Instead of the two broad excitation bands observed in the data (Fig.~\ref{fig:diffuse_scattering}), three sharp dispersionless modes are predicted by spin-wave theory. While two of these modes are gapped at 0.48 and 1.2 meV, the third lies at elastic energies (Fig.~\ref{fig:osiris}). In order to make a more detailed comparison to the data, these spin-wave modes are artificially broadened by a FWHM of 0.4 meV. Figure~\ref{fig:diffuse_scattering}(e) shows a constant energy slice through the spin-wave simulation at 0.25 meV in the ($h$,$k$,0)-plane, which captures scattering from both the 0.48 meV and the elastic modes due to this broadening. Comparison of this slice to the data in Fig.~\ref{fig:diffuse_scattering}(b) shows that it is able to reproduce some of the ring-like features although of course these features are much sharper in the simulation. Similar qualitative agreement is achieved by the spin-wave slice at 0.9 meV which captures the upper spin-wave mode (Fig.~\ref{fig:diffuse_scattering}(h)) and reproduces the broad blocks of scattering in the data. As expected a slice taken perpendicular to the kagome layers at 0.25 meV shows dispersionless streaks of scattering parallel to the $l$-direction in agreement with the observed 2D nature of the magnetism.

\section{Discussion}

\begin{figure}
\includegraphics[width=\columnwidth]{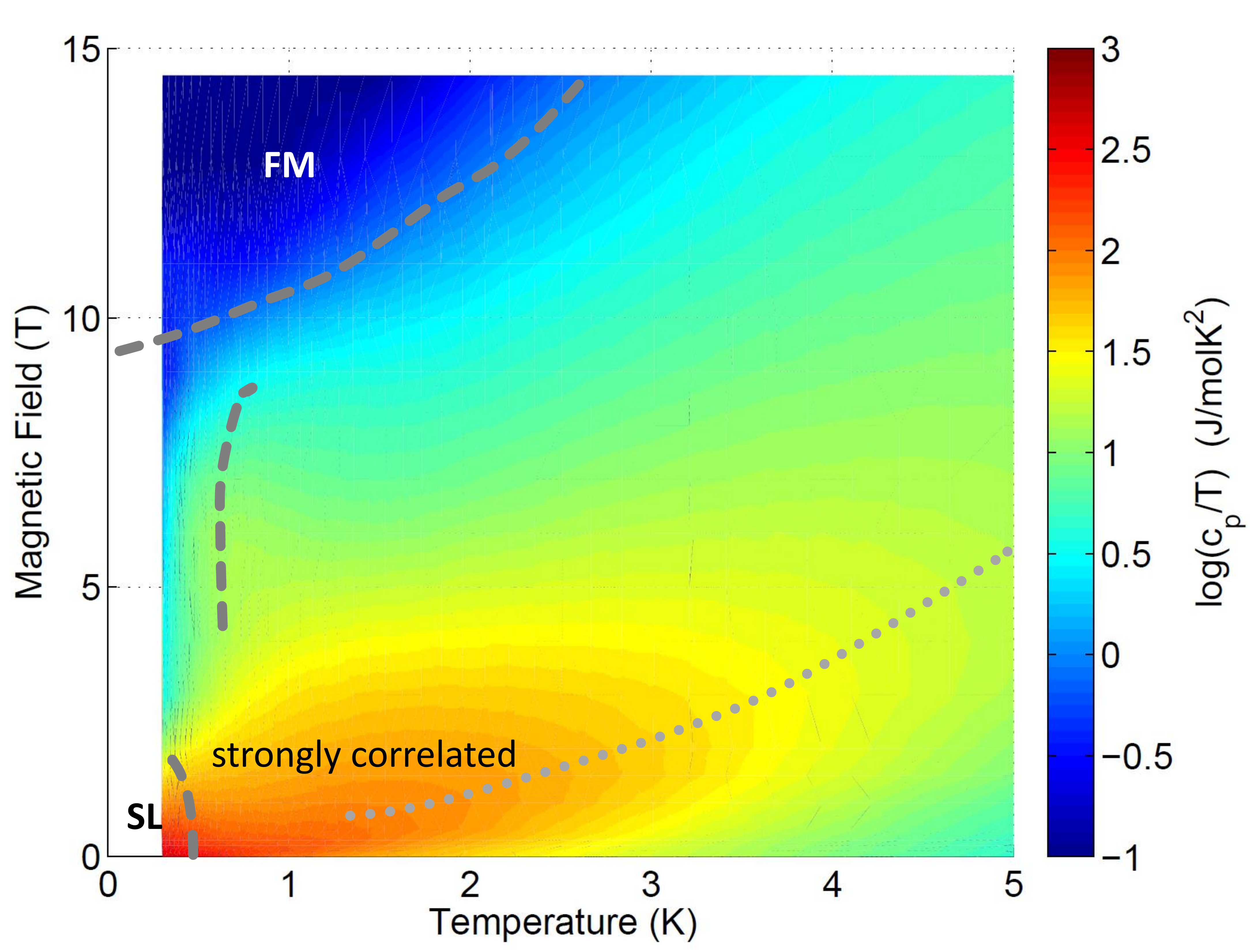}
\caption{\textbf{Proposed magnetic phase diagram plotted over the magnetic heat capacity.} The color plot gives $\log(C_p/T)$ versus temperature and magnetic field. Proposed crossovers are indicated by the dashed lines while the dotted line indicates the maximum of the Schottky-like peak. The phases are labeled as spin liquid (SL) at low fields and temperatures, strongly correlated precursor spin liquid at higher temperatures and ferromagnet (FM) at highest fields.
\label{fig:phase_dia}}
\end{figure}

One of the main questions raised by this investigation is why the Hamiltonian of \cacro supports a quantum spin liquid ground state. As we have shown here, the $S=\nicefrac{1}{2}$ \Cr ions form bilayers where both layers that make up the bilayer are distorted kagome planes made from strong ferromagnetically coupled equilateral triangles alternating with weaker antiferromagnetically coupled equilateral triangles. These two planes have similar but not identical interaction strengths and are coupled together by a weak ferromagnetic interaction so that the ferromagnetic triangles in one plane couple directly to the antiferromagnetic triangles in the other plane and vice versa. All the interactions are Heisenberg and there is no discernible anisotropy. 

Ferromagnetic interactions are a well-known ingredient of spin liquids that have strong anisotropy if this anisotropy favors a non-collinear spin arrangement that competes with the parallel alignment favored by the interactions. The best example is spin ice where the magnetic ions form a pyrochlore lattice (corner-sharing tetrahedra) and interact via ferromagnetic dipolar interactions, which compete with the anisotropy that forces each spin to point along its local [1,1,1] direction \cite{Gin14}. For magnets with isotropic interactions and no anisotropy the frustration must be generated entirely by competition between the interactions. In this situation it is generally assumed that the interactions should be predominantly antiferromagnetic as for the kagome spin liquid realized by Herbertsmithite. This is because for isotropic ferromagnetic interactions, the spins can rotate and align parallel to each other on any lattice giving rise to ferromagnetic long-range order. An example of an isotropic spin liquid that has ferromagnetic interactions is Kapellasite where the $S-\nicefrac{1}{2}$ Cu$^{2+}$ ions form a kagome lattice and interact via ferromagnetic first neighbor interactions which compete with antiferromagnetic third neighbor interactions (diagonally across the hexagon) \cite{Fak12}. These two interactions have approximately equal strength and the system can be regarded as a network of $S-\nicefrac{1}{2}$ Heisenberg antiferromagnetic chains with frustrated interchain coupling \cite{Gon16}. 

In order to gain further insight into the origins of the frustration in \cacro and the role played by the ferromagnetic interactions, the classical ground state from the kagome bilayers model was calculated from mean-field theory. It should of course be emphasized that this is not the real ground state which is characterized by dynamic spin motion and the absence of long-range magnetic order down to the lowest temperatures as shown in Ref.~\cite{Bal16}. Nevertheless the classical ground state can provide us with a snapshot of the spin arrangement. As shown in Fig.~\ref{fig:struct}(e) the direct (vertex to vertex) coupling of a FM triangle (upper triangle) to an AFM triangle (lower triangle) yields a motif of two coupled opposite-sign triangles whose ground states are clearly incompatible with each other. The coupling which is due to the weak FM interaction J0 prevents either triangle from achieving its ideal ground state of collinear order on the FM triangle and 120$^\circ$ order in the AFM triangle. This frustration introduces quantum fluctuations which destabilize long-range order, destroy the spin-wave excitations and give rise to the spin liquid state in \cacro.

While spin-wave theory clearly cannot account for the broad, diffuse magnetic excitation spectrum observed in zero magnetic field, it does provide a good description of the spectrum observed with an applied field of $H=11$~T which is close to the saturation field where the ground state has long-range ferromagnetic order. Here sharp spin-wave branches are observed which are gapped and dispersive. The agreement however between spin-wave theory and the data becomes increasing worse as the field is reduced. The well-defined spin-wave modes observed in \cacro broaden and lose their dispersion and the gap becomes smaller and less clean. At $H=1$~T,  the lowest energy excitation branch appears to touch the elastic line and at zero field the spectrum is gapless and consists of two broad and diffuse dispersionless bands. 

The heat capacity and magnetization provide additional insight. Heat capacity reveals the absence of any phase transition in \cacro as a function of applied field and temperature, nevertheless it shows several interesting features. In particular a peak in $C_p/T$ appears at $H=1$~T and $T=0.4$~K which is the same field where the lowest magnetic excitation branch first starts to emerge from the ground state in the inelastic neutron data and where the magnetization shows a change of slope. Together these results suggest that at $H=1$~T \cacro crosses over from its spin liquid ground state to another as yet unknown regime. The  boundary of the spin liquid phase as a function of temperature is probably $T=0.4$~K where the peak in $C_p/T$ is observed. Below this temperature the spins enter a regime of persistent slow dynamics as revealed by muon spectroscopy in Ref.~\cite{Bal16}. The phase at higher temperatures can be described as a strongly correlated precursor spin liquid since the spectrum measured by neutron scattering shows a similar diffuse spectrum at $T=1.6$~K as at $T=0.09$~K. 

Figure \ref{fig:phase_dia} shows the proposed phase diagram as a function of magnetic field and temperature plotted over the magnetic heat capacity ($\log[C_p/T]$). Beside the spin liquid phase, the strongly correlated precursor phase and of course the ferromagnetic phase at highest fields, another crossover was observed for fields of $H=6-10$~T and temperatures of $T\approx0.7$~K which is yet to be explored. 

\section{Conclusion}

To conclude, the quantum spin liquid ground state of \cacro arises from a complex combination of FM and AFM Heisenberg interactions. High-field single crystal inelastic neutron scattering allowed us to measure the exchange couplings directly, providing an accurate model of the Hamiltonian. This makes \cacro one of the rare examples of a quantum spin liquid for which the Hamiltonian is quantitatively known and reveals that the frustration arises from a previously unexplored mechanism due to the direct coupling of FM triangles to AFM triangles. The magnetic Hamiltonian is used to analyze our experimental findings as a function of magnetic field and a consistent picture of a spin liquid ground state in zero field with crossovers to several other phases as a function of field and temperature is achieved. A theoretical model of the ground states and excitations under zero- and intermediate fields is highly desired and will be pursued by the authors in the near future.

\begin{acknowledgments}
\emph{Acknowledgments} - We thank S. Toth for his help with the spin wave fitting and J. Reuther for helpful discussions. This work utilized facilities supported in part by the National Science Foundation under Agreement No. DMR-1508249.
\end{acknowledgments}

\appendix

\section{APPENDIX A: Further experimental details}

Powder and single crystal samples of \cacro were prepared according to the synthesis route described in Ref. \cite{Bal16_2}. DC magnetic susceptibility was measured from 400~K down to 1.8 K on a 49~mg single crystal sample using a Quantum Design MPMS Squid sensor both along the a and c axis. The applied DC field was 0.1~T. Magnetization up to 14 T was measured on 15~mg sngle crystal using a Quantum Design PPMS system at 1.8 and 3 K. The heat capacity was measured down to 300 mK on a 0.91 mg single crystal using a relaxation technique. External magnetic fields of up to 14.5 T were applied both parallel and perpendicular to the $c$ axis. These measurements were performed at the Laboratory for Magnetic Measurements, Helmholtz Zentrum Berlin f\"ur Materialien und Energie, Germany.

Powder inelastic neutron scattering was measured at the time-of-flight spectrometer TOFTOF at the Heinz Maier Leibnitz Zentrum, Munich, Germany on a 8.14 g sample and at 430 mK. An incident energy of 3.27 meV was used, giving a resolution of 0.08 meV. The data were binned into steps of 0.02 {\AA} and 0.02 meV. Single crystal inelastic neutron scattering was measured at the triple axis spectrometers IN14 (Institut Laue-Langevin, Grenoble, France) and MACS II (NIST Center for Neutron Research, Gaithersburg, USA) as well as on the time-of-flight spectrometers LET and OSIRIS (both ISIS facility, Didcot, UK). For all experiments two co-aligned rod shaped single crystals with a total mass of 1.71 g and a mosaicity of less than 2$^{\circ}$ were used. At IN14 the flatcone option was used and the scattering planes $h0l$ and $hk0$ were scanned. The measurements took place at a temperature of 1.6 K and the final neutron energy was as fixed to 4.06 meV, giving a resolution of 0.12 meV. The data was binned into steps of 0.0085 along $[h,h,0]$, 0.013 along $[h,-h,0]$ and 0.06 along $[0,0,l]$, and the binned slices were then smoothed by a weighted average over 1.5 bins. For the MACS II experiment the $hk0$ plane was horizontal and the temperature was at 90 mK. A vertical magnetic field of 11~T was applied parallel to [0,0,L]. The final energy was fixed to 3.7 meV and the incident energy was varied between 4.5-6.7~meV giving a resolution of 0.35-0.65~meV. The data was binned into steps of 0.075 along $[h,h,0]$ and $[h,-h,0]$ and combined into a 3D dataset using the DAVE software package \cite{Azu09}. At LET we used $E_i=3.63$~meV with $\Delta E=0.3$~meV to obtain 3D dataset at 2~K and 9~T parallel to [0,0,L]. The sample was rotated for 60$^{\circ}$ in 2$^{\circ}$ steps. The OSIRIS measurement was performed in the $hk0$ plane at a temperature of 220 mK. The magnetic field was varied between 0-7.5~T again applied parallel to [0,0,L]. Energy transfers between -0.05 and 2~meV were measured with a resolution at the elastic line of 0.025~meV. The data was binned into steps of 0.05 r.l.u. and 0.05 meV, and smoothed with a hat function
of width 2 bins. For both LET and OSIRIS data we used the Horace software package for visualization \cite{Ewi16}.

\section{APPENDIX B: Heat capacity data treatment}

\begin{figure}
\includegraphics[width=\columnwidth]{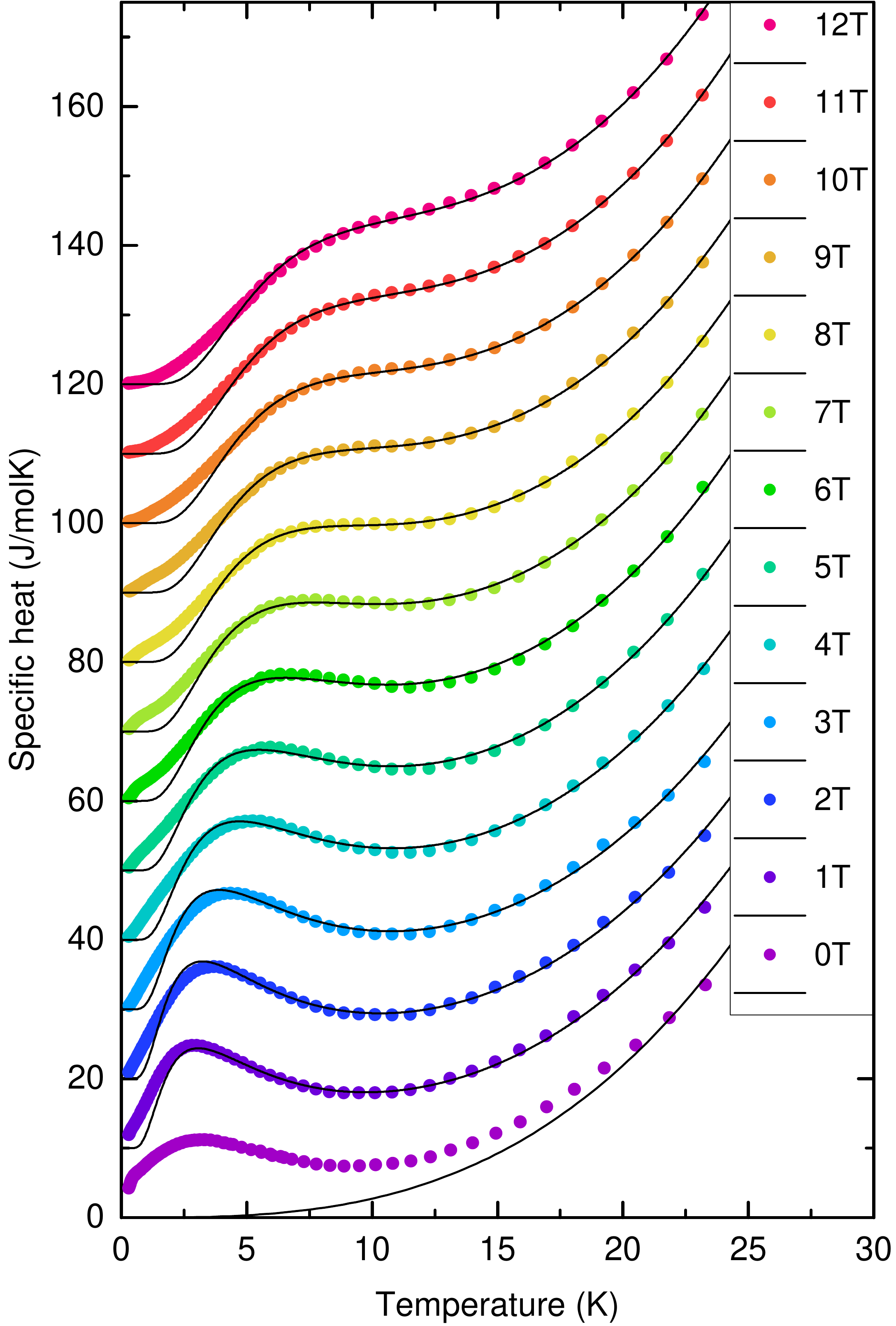}
\caption{APPENDIX B: Raw heat capacity data together with fits to the sum of the phonon and Schottky contributions above 4~K described in the text.}\label{fig:heat_app}
\end{figure}

The heat capacity under external magnetic field consists of several contributions. First of all it contains a field-independent Debye-like phonon contribution $C_{\text{Debye}}\sim T^3$. It also contains a Schottky-type anomaly which can be described as a two-level system whose gap $\Delta$ is field dependent due to Zeeman splitting
\begin{align}
C_{\text{Schottky}}\sim \left(\frac{\Delta}{T}\right)^2\cdot e^{(-\Delta/T)}
\end{align}
Furthermore there are additional magnetic terms in the heat capacity which become important at low temperatures and not too high fields. The weighted sum of the phonon and Schottky contributions, $A\cdot C_{\text{Schottky}}+B\cdot C_{\text{Debye}}$, was fitted to data points above $T=4$~K for measurements under external field simultaneously as shown in Fig.~\ref{fig:heat_app}. From these fits the phonon contribution could be obtained which is plotted over the zero field measurement (solid line). The phonon contribution was subtracted from all heat capacity data shown in figures~\ref{fig:bulk}(c), \ref{fig:cp}, and \ref{fig:phase_dia}.

\section{APPENDIX C: Details of linear spin wave theory analysis}\label{App:LSWT}

\begin{figure*}
\includegraphics[width=0.65\textwidth]{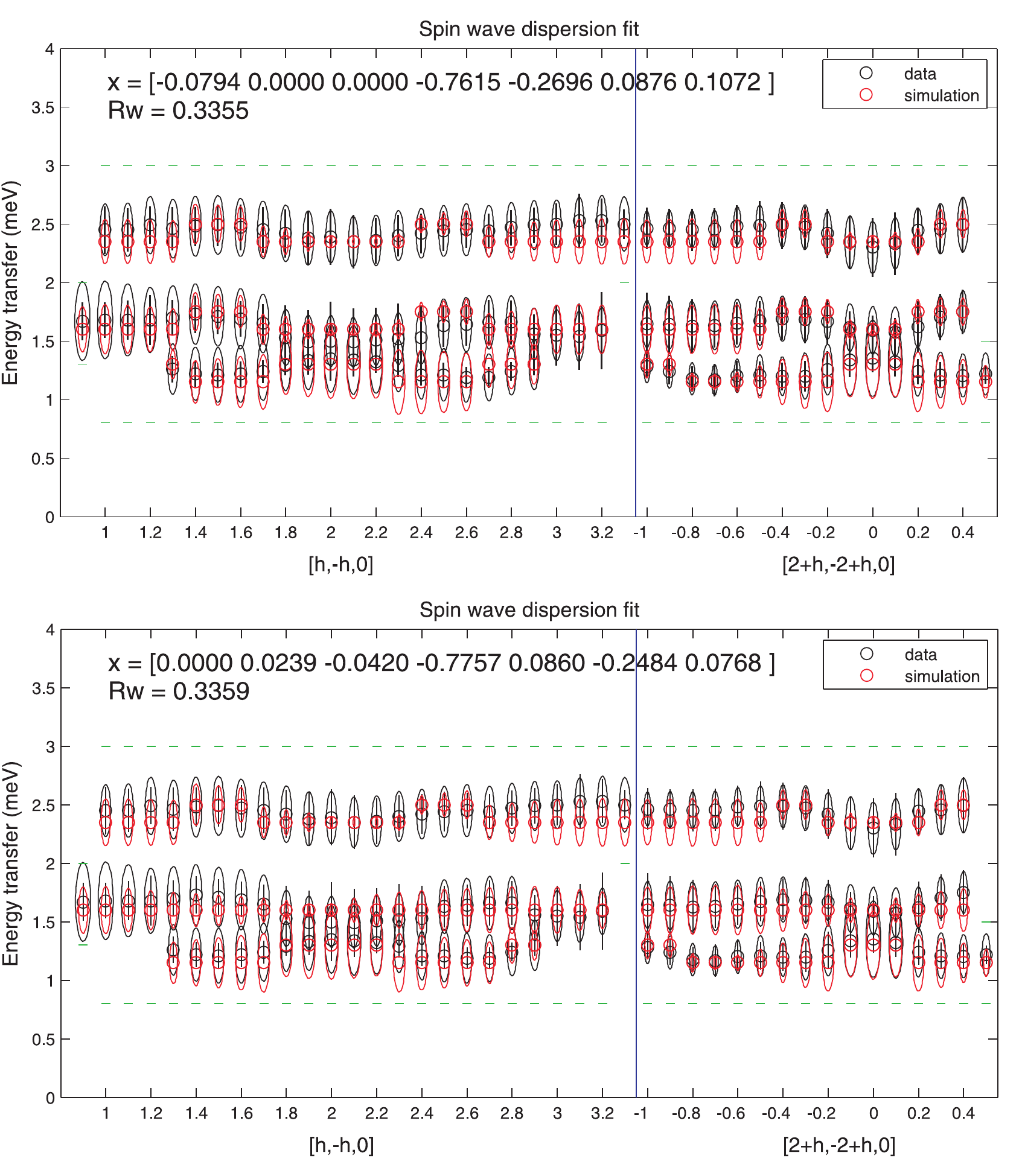}
\caption{APPENDIX C: Graphical representation of the best fit of the kagome bilayer (top) and the coupled hexagons model (bottom) to the data. Different directions in $Q$ space are separated by the vertical blue line. The data from MACS is presented in black and the fit result in red. The circles show the center of the spin wave modes and the ellipsoids indicate their intensity. In addition, the width of the modes in the data is given by the vertical black lines. The boundaries of the energy scans are shown in green. The fitted exchange constants are $x=[J0,J11,J12,J21,J22,J31,J32]$ and Rw is the weighted R-Factor of the fit.}\label{fig:fit_res}
\end{figure*}

\begin{figure*}
\includegraphics[width=0.8\textwidth]{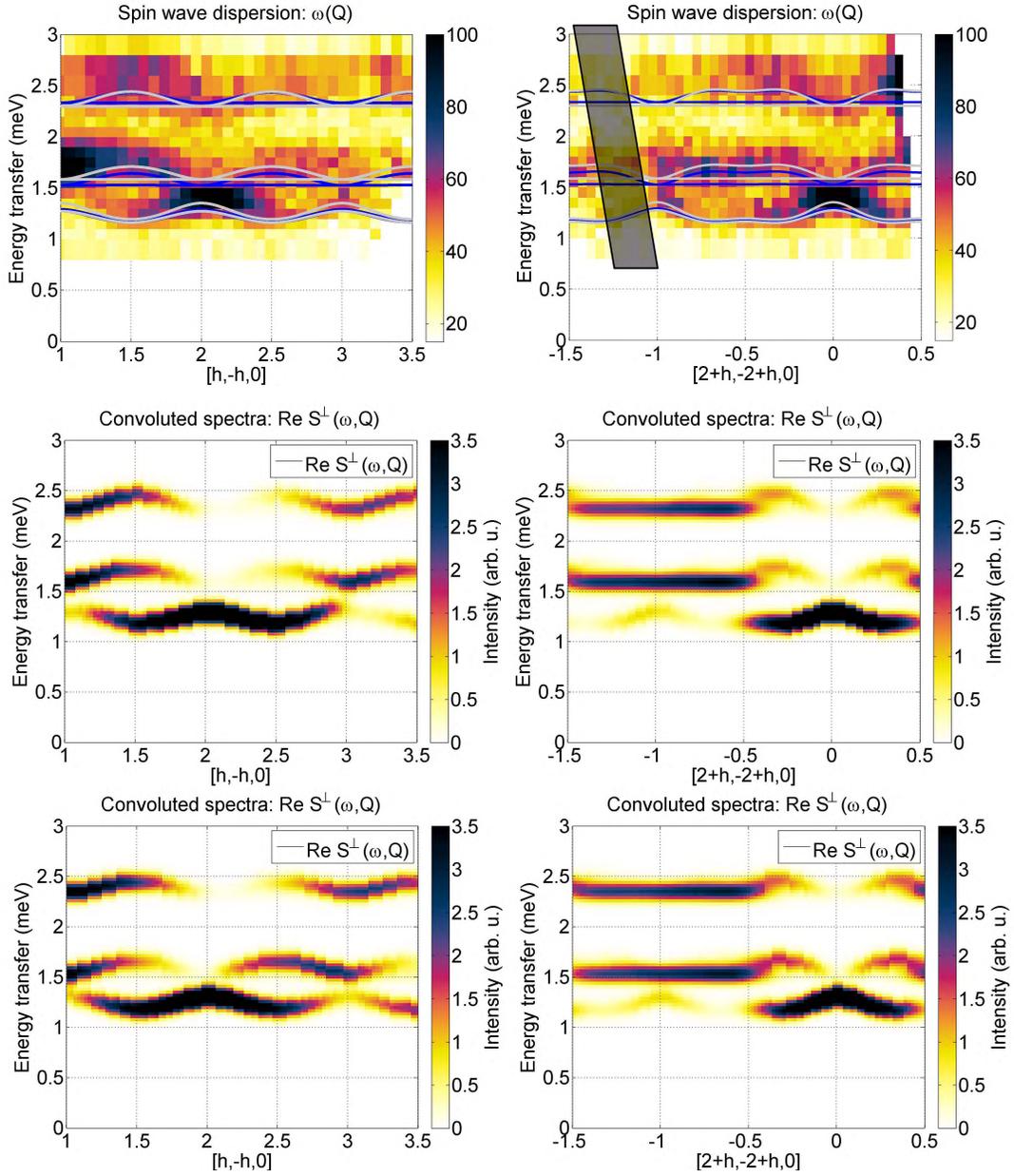}
\caption{APPENDIX C: Comparison of the kagome bilayer and the hexagon model using INS data at 11~T. (first row) INS data from MACS at $H=11$~T and $T=90$~mK. The dispersions obtained from the fitting procedure are plotted over the data: Grey line = kagome bilayer model, blue line = hexagon model. The shaded region indicates the dark angle of the magnet. The second and third row show the calculated spectra convolved with a resolution of 0.4~meV giving the simulated intensity distribution for the kagome bilayer (second row) and the hexagon model (third row) respectively.}\label{fig:MACS_comparison}
\end{figure*}

In order to fit the data to linear spin-wave theory, the spin-wave dispersions were first extracted from the data measured on the MACS instrument at 11T. Constant-wavevector cuts were taken through the two principal dispersion directions shown in Fig.~\ref{fig:spin_waves_11T}(a) and (c) at every 0.1 step in r.l.u., and the obtained intensity versus energy cuts were fitted to a sum of Gaussians functions and a constant background. The size of the energy step was given by the interval between the measured constant energy slices (0.1-0.2~meV). From these fits the energy positions, widths and the intensity of the spin wave modes were obtained as shown in Fig.~\ref{fig:fit_res}.

The fitting procedure itself was performed as follows: The Hamiltonian was set up according to the respective model (single kagome layer, kagome bilayer or coupled hexagons) and random starting parameters were generated for the $J$s within the constraints of $|J_{ij}|\leq1.4$~meV. The Hamiltonian was then diagonalized and a total of 6 spin-wave modes were found. This is expected since the number of spin-wave excitations equals the number of magnetic atoms in the unit cell which is 6 for the 2D magnetic structures of Ca$_{10}$Cr$_7$O$_{28}$ under consideration. It should be noted that since a maximum of 3 modes are observed in the data some of the spin-wave branches must be closely spaced or have no intensity. The energies and neutron scattering cross-sections for each mode were then calculated at every wavevector. In order to compare them to the data the modes were binned in energy steps of 0.15~meV which corresponds to the minimum observed FWHM in the data. The spin wave intensity from the simulation was used to select the bins with the largest intensity which were kept up to the number of modes observed in the data at the respective $\bm{Q}$ position (maximum 3 for Ca$_{10}$Cr$_7$O$_{28}$) and the remaining ones were discarded.

The modes were then sorted according to their energy. The goodness of fit was calculated for each wavevector from the difference between energies of the observed peaks and position of the calculated energy bins. The goodness of fit values for each mode were weighted by the experimentally determined inverse width in energy of that mode, where the sharpest mode was normalized to have weight=1. In this way more weight was given to peaks which were better defined. A total of 40 constant wavevector cuts extracted from the slices along $[h,-h,0]$ and $[2+h,-2+h,0]$ were fitted simultaneously. The goodness of fit values were then summed over all wavevectors and modes and their sum was minimized using a simplex algorithm provided with Matlab allowing all $J$s of the model to vary within the constraints. Having reached a minimum, the values for the $J$s were saved together with the corresponding weighted R-Factor. 

The same procedure was repeated for the next set of starting parameters. In this way approximately 1000 minimization runs with random starting parameters could be performed in 12 hours on a regular desktop computer. The best fits were sorted according to their weighted R-Factors and were simulated to make a visual comparison with other dispersion directions and also the 9T data collected on LET. The single kagome layer model was unable to fit the data adequately, however good fits were achieved for both the kagome bilayer and the coupled hexagons model (see Fig.~\ref{fig:MACS_comparison}).

\end{document}